# DNA-decorated graphene chemical sensors

*Ye Lu, Brett R. Goldsmith, Nicholas J. Kybert, A.T. Charlie Johnson\**

*University of Pennsylvania, Department of Physics and Astronomy, 209 S. 33rd St., Philadelphia PA 19104-6396*
RECEIVED DATE (automatically inserted by publisher); cjohnson@physics.upenn.edu

**Abstract**
Graphene is a true two dimensional material with exceptional electronic properties and enormous potential for practical applications. Graphene's promise as a chemical sensor material has been noted but there has been relatively little work on practical chemical sensing using graphene, and in particular how chemical functionalization may be used to sensitize graphene to chemical vapors. Here we show one route towards improving the ability of graphene to work as a chemical sensor by using single stranded DNA as a sensitizing agent. The resulting broad response devices show fast response times, complete and rapid recovery to baseline at room temperature, and discrimination between several similar vapor analytes.

Graphene has been actively studied as a chemical sensor since shortly after it was isolated in 2004 [1-3]. Increasingly sophisticated device processing has revealed that early measurements of graphene exhibited chemical sensing responses that were amplified by unintentional functionalization [4]. Here, we start for the first time with chemically clean graphene transistors that are inert to a variety of chemical vapors. We then purposefully functionalize the graphene to generate devices with different chemical sensing responses. We demonstrate that graphene can be combined with ssDNA to create a chemically diverse family of vapor sensors that is promising for use in a "nose-like" vapor sensing system.

Nose-like sensing schemes derive their organizational principle from biological olfactory systems, where a relatively small number (100s) of sensor types are deployed with broad and overlapping sensitivities to a much larger number of volatile analytes [5,6]. In our DNA-graphene sensor system, ssDNA is not used for its biological functionality, but instead provides *sequence-dependent* chemical recognition capability, potentially enabling the required number (hundreds) of chemically distinct sensor responses. Reduced graphene oxide, or "chemically derived graphene", has also shown potential as a vapor sensor material where residual oxygen defects (e.g., carboxylic acids or epoxides) provide binding sites for analyte molecules.[7]

Graphene transistors were constructed using exfoliated kish graphite on silicon substrates with a 300 nm oxide layer [4]. Devices were carefully cleaned to prevent spurious sensing results [4,8], then functionalized with a self-assembled layer of ssDNA as done previously for carbon nanotube devices [9,10]. Two ssDNA sequences ('Sequence 1' and 'Sequence 2') were selected because of their prior use in vapor sensors based on electronic[9] and optical fluorescence[11] readout strategies.

Sequence 1: 5' GAG TCT GTG GAG GAG GTA GTC 3'

Sequence 2: 5' CTT CTG TCT TGA TGT TTG TCA AAC 3'

AFM measurements showed that the self-assembled ssDNA layer had a thickness of approximately 0.5 nm (Fig. 1a). Although ssDNA films on graphene did not have visible holes or aggregates, AFM revealed an RMS roughness of 0.4 nm, about twice as large as that of pristine graphene. We did not observe ssDNA deposition on the $SiO_2$, nor did the roughness of that surface change.

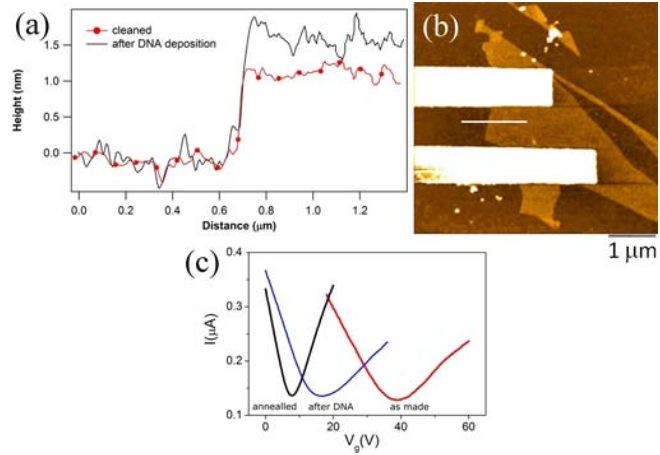

**Figure 1** (a) AFM line scans of ssDNA on graphene. (b) AFM image with z-scale of 25 nm. White lines indicate the scan lines of Fig. 1a. (c) I-$V_G$ characteristics for a graphene device through the steps of functionalization showing the expected doping shifts due to ssDNA application.

Figure 1b shows how the current-gate voltage (I-$V_G$) characteristic of an individual device changed as the graphene was cleaned and then chemically modified. Carrier mobility and doped carrier densities are extracted from such data as in Ref. [12]. For the as-fabricated device, the hole and electron mobilities were 1000 cm$^2$/V-s and 750 cm$^2$/V-s, respectively. The doped charge carrier density (carrier density at $V_G$=0) was 3.3×10$^{12}$/cm$^2$ holes. After the cleaning process, both the hole and electron mobility increased to 2600 cm$^2$/V-s, and the doped carrier density decreased to 6.2×10$^{11}$/cm$^2$ holes. After functionalization with ssDNA, the hole and electron mobilities decreased to 1600 cm$^2$/V-s and 750 cm$^2$/V-s, respectively, indicating slightly increased carrier scattering due to ssDNA on the graphene surface [13]. The doped charge carrier density with the ssDNA layer was 1.4×10$^{12}$/cm$^2$.

Application of ssDNA led to a shift in the I-$V_G$ minimum indicating an increase in hole density (Fig. 1b). The polarity of this shift is consistent with chemical gating by negatively charged molecules in the vicinity of the graphene. Using computer models of ssDNA on carbon nanotubes, we estimate an adsorption density of ~ 1.5×10$^{14}$ bases/cm$^2$ at 100% coverage [14]. Even for very weak electrostatic interactions of ssDNA bases with graphene, such a large number of negatively charged bases would easily account for the observed shift in the electrostatic doping of $6.2 \times 10^{11}$ /cm$^2$ holes.

Chemical sensing experiments were performed in a controlled environmental chamber. The device current was monitored while applying a 1 mV bias voltage and zero gate voltage. Initially, nitrogen carrier gas was flowed through the chamber at a rate of 1 sLm. Analyte gases were substituted for a small percentage of the nitrogen flow with the total flow rate held constant. In order to compare changes in response, the data are presented as changes in current normalized to the device current measured in a pure Ar flow.

In Figure 2, we compare responses to vapors of devices based on clean graphene, graphene functionalized with ssDNA Seq. 1, and graphene functionalized with Seq. 2. The vapors used in Fig 2a-b were dimethylmethylphosphonate (DMMP) and propionic acid, respectively. For both analytes, the current response of clean graphene was very low and barely detectable above system noise, although a response $\Delta I/I_0 \sim 1\%$ was observed at the highest concentrations tested. After coating with ssDNA, enhanced responses on the scale of 5-50% were observed. Responses were reproducible, with nearly perfect recovery to baseline upon purging. As was suggested for ssDNA-nanotube devices,[9] we infer that the role of the ssDNA is to concentrate water and analyte molecules near the otherwise chemically inert and hydrophobic conduction channel, and in this way greatly increase the current response compared to that of bare graphene. For these two analytes the sign of the current responses were consistent with a chemical gating effect on the graphene channel where hole conduction dominates. DMMP, a strong electron donor [15], is expected to become positively charged, consistent with decreased device current. Conversely, propionic acid is expected to donate a proton to residual water and acquire a negative charge, increasing device current.

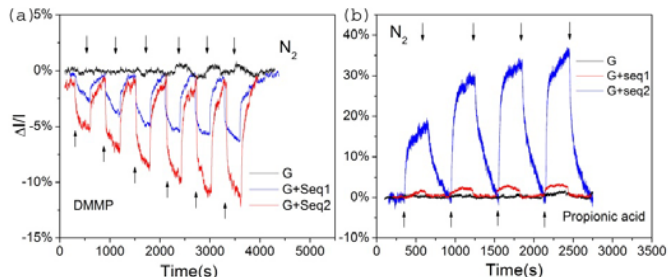

**Figure 2** Normalized changes in current versus time for ssDNA-graphene vapor responses. Lower arrows indicate introduction of analyte at progressively larger concentrations, while upper arrows indicate flushing with pure carrier gas. Clean graphene devices (black data) show very weak vapor responses that are barely above the noise floor. Devices functionalized with Seq. 1 (red data) or Seq. 2 (blue data) show significant responses that are sequence-dependent. (a) Measurement of DMMP at concentrations of 20, 40, 60, 80, 100 and 120 ppm (b) Measurement of propionic acid at concentrations of 90, 220, 435 and 650 ppm.

For both analytes in Figure 2, sensing response and recovery to baseline typically showed two distinct timescales. The initial response occurred within a fraction of a second, while the slower equilibration took up to several hundred seconds. This has been observed for other sensor types and may indicate the presence of a two-stage molecular binding process [16].

**Table 1.** Sensing Response ($\Delta I/I_0$) for several analytes.

| odor | conc (ppm) | pristine graphene | graphene + Seq 1 | graphene + Seq 2 |
|---|---|---|---|---|
| DMMP | 20 | <0.1% | -2.5% | -5% |
| Prop Acid | 90 | <0.1% | 1.5% | 18% |
| Methanol | 7500 | 0.5% | 1% | 2% |
| Octanal | 14 | 0.5% | 1% | 3% |
| Nonanal | 0.6 | 0.5% | 2% | 8% |
| Decanal | 1.7 | 4% | 2% | 4% |

Six different analyte responses are shown in Table 1. Three compounds are homologous aldehydes, with linear chemical formulas $CH_3(CH_2)_NCHO$, where N=5, 6, 7. The significantly different current responses seen for this sequence shows a chemical differentiation in an atmospheric sensor not often seen outside of biological systems. This may reflect a complex sensing mechanism beyond the simple charging mechanism proposed above. Molecular dynamics simulations of ssDNA adsorbed onto carbon nanotubes suggests complex conformational motifs may influence the chemical affinity of the device [17]. Similar effects may occur for ssDNA-graphene, leading to a sensing response that is the result of a combination of analyte-DNA and analyte-graphene interactions. Clean graphene devices exhibit very small responses to all analytes at all concentrations tested, with the exception of decanal.

The ssDNA functionalization procedure does not seem to increase the noise power of the device. The noise amplitude was measured by taking a power spectrum of a typical background current when the device is exposed to nitrogen, and reading off the amplitude at 1Hz. This gives a 1/f noise amplitude of $6.25 \times 10^{-6}$ $A^2$/Hz, normalized for the 500 nA current to $2.5 \times 10^{-6}$ 1/Hz, within the typical current-normalized range for single layer graphene[18]. Converting the current noise power density to current variation yields a best-possible detection threshold of ~ 0.1% of the baseline current using a 1 Hz bandwidth. Assuming a linear response to DMMP concentration, this implies a detection limit of 0.4 ppm for a single device for DMMP.

Future experiments could evaluate the utility of large graphene sensor arrays for "electronic nose" systems. Recent developments in large scale synthesis of graphene make it possible to attempt very large scale device integration, and may open graphene devices to applications unavailable to carbon nanotube devices[19].

**Acknowledgements** This work was supported by the Nano/Bio Interface Center through the National Science Foundation NSEC DMR-0425780 and an Intelligence Community Postdoctoral Fellowship (B.R.G; National Geospatial Agency Grant No. HM1582-07-1-2014). N.J.K recognizes the support the REU program of the Laboratory for Research on the Structure of Matter, NSF MRSEC DMR05-20020.

(1) Ao, Z. M.; Yang, J.; Li, S.; Jiang, Q. *Chem Phys Lett* **2008**, *461*, 276-279.
(2) Schedin, F.; Geim, A. K.; Morozov, S. V.; Hill, E. W.; Blake, P.; Katsnelson, M. I.; Novoselov, K. S. *Nat Mater* **2007**, *6*, 652-655.
(3) Zhang, Y. H.; Chen, Y. B.; Zhou, K. G.; Liu, C. H.; Zeng, J.; Zhang, H. L.; Peng, Y. *Nanotechnology* **2009**, *20*, 185504.
(4) Dan, Y. P.; Lu, Y.; Kybert, N. J.; Luo, Z. T.; Johnson, A. T. C. *Nano Lett* **2009**, *9*, 1472-1475.
(5) Hopfield, J. J. *Proc Natl Acad Sci USA* **1999**, *96*, 12506-12511.
(6) Shepherd, G. M. *PLoS Biology* **2004**, *2*, 572-575.
(7) Robinson, J. T.; Perkins, F. K.; Snow, E. S.; Wei, Z.; Sheehan, P. E. *Nano Letters* **2008**, 10.1021/nl8013007.
(8) Ishigami, M.; Chen, J. H.; Cullen, W. G.; Fuhrer, M. S.; Williams, E. D. *Nano Letters* **2007**, *7*, 1643.
(9) Staii, C.; Chen, M.; Gelperin, A.; Johnson, A. T. *Nano Letters* **2005**, *5*, 1774 - 1778.
(10) See supplementary material at [URL will be inserted by AIP] for details of procedures of graphene transistor fabrication, cleaning, and functionalization.
(11) White, J.; Truesdell, K.; Williams, L. B.; AtKisson, M. S.; Kauer, J. S. *Plos Biology* **2008**, *6*, 30-36.
(12) Chen, J. H.; Jang, C.; Xiao, S. D.; Ishigami, M.; Fuhrer, M. S. *Nature Nanotechnology* **2008**, *3*, 206-209.
(13) Seiyama, T.; Kato, A.; Fujiishi, K.; Nagatani, M. *Anal Chem* **1962**, *34*, 1502-&.
(14) Johnson, R. R.; Johnson, A. T. C.; Klein, M. L. *Nano Lett* **2008**, *8*, 69-75.
(15) Novak, J. P.; Snow, E. S.; Houser, E. J.; Park, D.; Stepnowski, J. L.; McGill, R. A. *Applied Physics Letters* **2003**, *83*, 4026-4028.
(16) Albert, K. J.; Lewis, N. S.; Schauer, C. L.; Sotzing, G. A.; Stitzel, S. E.; Vaid, T. P.; Walt, D. R. *Chem Rev* **2000**, *100*, 2595-2626.
(17) Johnson, R. R.; Kohlmeyer, A.; Johnson, A. T. C.; Klein, M. L. *Nano Lett* **2009**, *9*, 537-541.
(18) Chen, Z. H.; Lin, Y. M.; Rooks, M. J.; Avouris, P. *Physica E* **2007**, *40*, 228-232.
(19) Li, X. S.; Cai, W. W.; An, J. H.; Kim, S.; Nah, J.; Yang, D. X.; Piner, R.; Velamakanni, A.; Jung, I.; Tutuc, E.; Banerjee, S. K.; Colombo, L.; Ruoff, R. S. *Science* **2009**, *324*, 1312-1314.

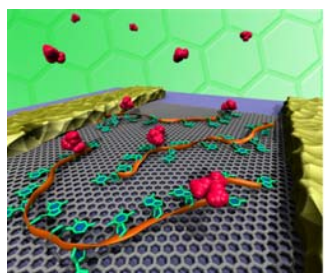
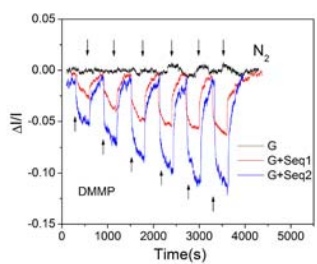